\documentclass[aps,prl,twocolumn,showpacs,superscriptaddress]{revtex4}
\usepackage{graphicx}

\begin{document}

\title{High Energy Spin Excitations in Electron-Doped Superconducting 
Pr$_{0.88}$LaCe$_{0.12}$CuO$_{4-\delta}$ with $T_c=21$ K}

\author{Stephen D. Wilson}
\affiliation{
Department of Physics and Astronomy, The University of Tennessee, Knoxville, Tennessee 37996-1200, USA
}
\author{Shiliang Li}
\affiliation{
Department of Physics and Astronomy, The University of Tennessee, Knoxville, Tennessee 37996-1200, USA
}
\author{H. Woo}
\affiliation{
Department of Physics and Astronomy, The University of Tennessee, Knoxville, Tennessee 37996-1200, USA
}
\author{Pengcheng Dai}
\email{daip@ornl.gov}
\affiliation{
Department of Physics and Astronomy, The University of Tennessee, Knoxville, Tennessee 37996-1200, USA
}
\affiliation{
Center for Neutron Scattering, Oak Ridge National Laboratory, Oak Ridge, Tennessee 37831-6393, USA}
\author{H. A. Mook}
\affiliation{
Center for Neutron Scattering, Oak Ridge National Laboratory, Oak Ridge, Tennessee 37831-6393, USA}
\author{C. D. Frost}
\affiliation{
ISIS Facility, Rutherford Appleton Laboratory, Oxon OX11 0QX, UK
}
\author{Seiki Komiya}
\affiliation{
Central Research Institute of Electric Power Industry, Komae, Tokyo 201-8511, Japan
}
\author{Yoichi Ando}
\affiliation{
Central Research Institute of Electric Power Industry, Komae, Tokyo 201-8511, Japan
}

\begin{abstract}
We use high-resolution inelastic neutron scattering to study the low-temperature magnetic excitations of  
electron-doped superconducting Pr$_{0.88}$LaCe$_{0.12}$CuO$_{4-\delta}$ ($T_c=21\pm 1$ K) 
over a wide energy range ($4$ meV$\le\hbar\omega\le 330$ meV).
The effect of electron-doping is to cause a wave vector ($Q$) broadening in the low-energy ($\hbar\omega\le 80$ meV) commensurate spin fluctuations at ($\pi$,$\pi$) 
and to suppress the intensity of spin-wave-like excitations at high
energies ($\hbar\omega\ge 100$ meV). This leads to a substantial redistribution in the spectrum of the local dynamical 
spin susceptibility $\chi^{\prime\prime}(\omega)$, and reveals a new energy scale similar to 
that of the lightly hole-doped YB$_2$Cu$_3$O$_{6.353}$ ($T_c=18$ K).
\end{abstract}

\pacs{74.72.Jt, 61.12.Ld, 75.25.+z}

\maketitle

High-transition-temperature (high-$T_c$) superconductivity in copper oxides 
occurs when sufficient holes or electrons are doped into the CuO$_2$ planes of 
their insulating antiferromagnetic (AF) 
parent compounds \cite{kastner}. Given the close proximity of AF order and
superconductivity, it is important to understand the evolution of magnetic excitations 
in the AF ordered parent insulators upon chemical doping to produce metals and superconductors, 
as spin fluctuations may play a crucial role 
in the mechanism of superconductivity \cite{scalapino}. 
For the undoped parent compounds, where AF order gives a diffraction peak at 
wave vector ${\bf Q}=(\pi,\pi)$ or $(0.5,0.5)$ (Fig. 1a),
spin waves at energies ($\hbar\omega$) below 60 meV found by neutron scattering 
show commensurate excitations around $(\pi,\pi)$ because of the 
large AF nearest neighbor exchange coupling ($J_1> 100$ meV, Figs. 1a,1c, and 3a)  \cite{hayden96,bourges97,coldea}.
Upon hole-doping to induce metallicity and superconductivity, 
the low-energy spin fluctuations of La$_{2-x}$(Sr,Br)$_x$CuO$_4$ (LSCO)
form a quartet of 
incommensurate peaks at wave vectors away from $(\pi,\pi)$ 
\cite{cheong,yamada98,tranquada,christensen} that may arise from the presence of static or
dynamic spin stripes \cite{kivelson}. For hole-doped 
YBa$_2$Cu$_3$O$_{6+x}$ (YBCO) with $x\geq 0.45$, the magnetic excitation spectra have 
a commensurate resonance at $(\pi,\pi)$ and incommensurate spin fluctuations similar to 
that of LSCO below this
resonance \cite{rossat,mook,fong,dai,bourges,hayden,stock}. 
For energies above the resonance, the excitations are spin-wave-like for $x\leq 0.5$ \cite{bourges,stock} and 
become a ``box-like'' continuum at $x=0.6$ \cite{hayden}.
In the extremely underdoped regime ($x=0.353$, $T_c=18$ K), Stock {\it et al.} \cite{stock1} showed that
spin fluctuations are commensurate around $(\pi,\pi)$ and have a damped spin resonance 
around 2 meV.

While the evolution of spin excitations in hole-doped superconductors has become increasingly clear, 
it is crucial to determine the evolution of spin excitations 
in electron-doped materials, as particle-hole symmetry is
an important ingredient of any theory purporting to explain the mechanism of high-$T_c$ superconductivity.
Unfortunately, due to the difficulty of growing large high-quality single crystals 
required for inelastic neutron scattering experiments, there exist only a few studies exploring the low-energy
($\hbar\omega\leq 10$ meV) spin dynamics in electron-doped materials such as 
Nd$_{1.85}$Ce$_{0.15}$CuO$_4$ (NCCO) \cite{yamada99,yamada}, and consequently the overall magnetic response of 
electron-doped materials remains largely unknown.
Recently, we began to systematically 
investigate the evolution of AF order and magnetic excitations 
as Pr$_{0.88}$LaCe$_{0.12}$CuO$_{4-\delta}$ (PLCCO) is transformed 
from the as-grown AF insulator 
into an optimally 
electron-doped superconductor ($T_c=25$ K) without static AF order 
through an annealing process with a minor oxygen content $\delta$ 
modification \cite{dai05,kang05,wilson05}. 
We chose first to study underdoped PLCCO ($T_c\sim 21$ K, $T_N\sim 40$ K) for two reasons. 
First, this material is between the as-grown AF insulator and optimally doped PLCCO
and therefore has a larger magnetic signal than that of the optimally doped PLCCO \cite{dai05,kang05,wilson05,fujita}.  
Second, Pr$^{3+}$ possesses a nonmagnetic singlet ground state in PLCCO, different from 
the Nd$^{3+}$ magnetic ground state in NCCO \cite{boothroyd}.

In this Letter, we report the results of inelastic neutron scattering measurements that probe the low-temperature 
($T=7$ K) dynamic spin response of electron-doped PLCCO ($T_c=21\pm 1$ K) for energies from $4$ meV to $330$ meV. We determine $Q$ and $\omega$ dependence of 
the generalized magnetic susceptibility $\chi^{\prime\prime}({\bf Q},\omega)$ \cite{hayden96}. 
We find that the effect of electron doping into the AF insulating PLCCO 
is to cause a wave vector broadening in the low-energy commensurate
magnetic excitations at $(\pi,\pi)$, consistent with that of the NCCO \cite{yamada99,yamada}. At high energies ($\hbar\omega\geq 100$ meV), the excitations are 
spin-wave-like rings, but with a dispersion steeper than
that of the undoped Pr$_2$CuO$_4$ \cite{bourges97} and with a significant reduction in the spectral weight of 
the local dynamical spin susceptibility $\chi^{\prime\prime}(\omega)$ \cite{hayden96}. 
A comparison of PLCCO and lightly doped YBCO ($x=0.353$) \cite{stock1} reveals that the 
energy scale for $\chi^{\prime\prime}(\omega)$ in both materials is at
$\sim$2 meV, lower than the $\sim$18 meV for the optimally doped LSCO \cite{christensen}.
  
\begin{figure}
\includegraphics[scale=.30]{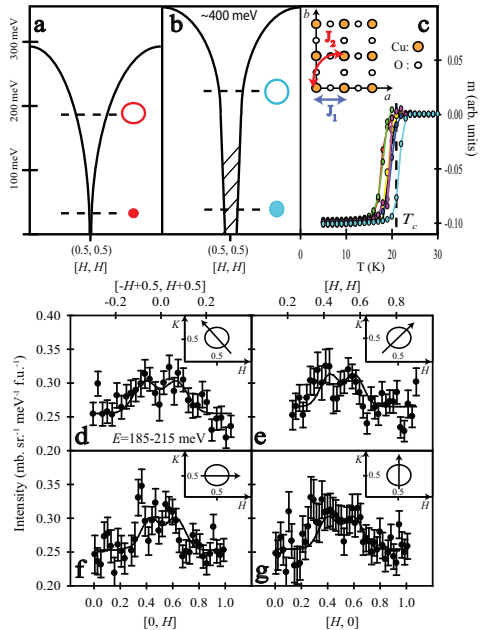}
\caption{
 Pictorial representation of the dispersions of the spin excitations in the a)  
 insulating Pr$_2$CuO$_4$ and b) superconducting PLCCO.  
 b) Magnetic susceptibility measurements of $
 T_c$'s in arbitrary units for the seven samples. The inset shows 
 unit cell with exchange couplings $J_1$ and $J_2$. d-g) 
 one-dimensional cuts through the spin excitations
 at $\hbar\omega=200\pm 15$ meV along the $[\overline{1},1]$, $[1,1]$,
 $[1,0]$, $[0,1]$ directions.
 }
\end{figure}

\begin{figure}
\includegraphics[scale=.30]{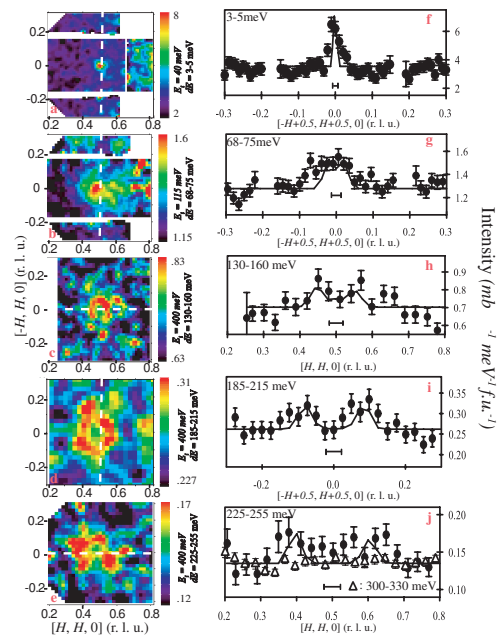}
\caption{$S({\bf Q},\omega)$ in the $(H,K)$ plane at a) $\hbar\omega=4\pm 1$, b) $71.5\pm 3.5$, c) $150\pm 10$, d) $200\pm 15$,
and e) $240\pm 15$ meV. The incident beam energy $E_i=40$, 115, 200, and 400 meV data have counting times of 18, 60, 44, and 76 hours respectively 
with a source proton current of 170 $\mu$A.
f-j) Narrow ${\bf Q}$-cuts passing through $(0.5,0.5)$ along the dashed line directions shown in a-e).
Upper open triangles in j) show a thicker cut at $\hbar\omega=315\pm 15$ meV along the $[1,1]$ direction (integrated from -0.2 to 0.2 along the $[\overline{1},1]$ direction).
Solid lines are the calculated one-magnon cross sections 
from the linear spin wave fit to the data with $J_1=162$ meV and $J_2=0$. Horizontal bars
below each cut show the instrumental resolution.}
\end{figure}

We grew seven high quality (mosaicity $<1^\circ$) PLCCO single crystals (with a total mass of
20.5 grams) using the traveling solvent method in a 
mirror image furnace. 
To obtain superconductivity, we annealed the
as-grown nonsuperconducting samples in a vacuum ($P< 10^{-6}$ mbar) at $T=765\pm 1$$^\circ$C for four days. While both ends of the same  
cylindrical shaped crystal were found to have identical $T_c$'s, 
there are small ($\pm1$ K) differences in $T_c$'s for separately annealed samples. Figure 1c shows
magnetic susceptibility measurements on all seven crystals used in our neutron experiments. They
have an average $T_c = 21\pm 1$ K.  For the experiment, we 
define the wave vector ${\bf Q}$ at $(q_x,q_y,q_z)$ as 
$(H,K,L)=(q_xa/2\pi,q_ya/2\pi,q_zc/2\pi)$ reciprocal lattice units (r.l.u) in
the tetragonal unit cell of PLCCO (space group $I4/mmm$, $a=3.98$, and $c=12.27$ \AA ).
The seven PLCCO crystals were co-aligned to within 1$^\circ$ in the $[H,H,L]$ zone  
using HB-1/HB-1A triple axis spectrometers at the High-Flux-Isotope reactor, Oak Ridge National 
Laboratory. Our inelastic neutron scattering experiments were 
performed on the MAPS time-of-flight spectrometer with the incident beam parallel to the $c$-axis 
($L$-direction) of PLCCO 
at the ISIS facility \cite{hayden}.
Four different incident beam energies of $E_i=40$, $115$, $200$, and $400$ meV were used,  
and the scattering was normalized to absolute units using a vanadium standard.

Figure 2 summarizes images of neutron scattering intensity $S({\bf Q},\omega)$ centered about $(\pi,\pi)$ at $T=7$ K in units of mbarns/sr/meV/f.u. without any background subtraction.  
At the lowest energy ($\hbar\omega=4\pm 1$ meV) probed (Fig. 2a), the scattering consists of a strong peak 
centered at $(\pi,\pi)$ with some phonon contamination evident at larger wave vectors. A constant-energy cut through the image reveals a commensurate peak on a flat background (Fig. 1f). 
The peak is significantly broader than
the instrumental resolution (horizontal bar) and gives a correlation length of $\sim$70 \AA. Upon increasing energy, the peak at $(\pi,\pi)$ broadens in width (Fig. 2b) and 
weakens in intensity (Fig. 2g). With further increase in energy to $\hbar\omega=145\pm 15$ meV, the scattering 
becomes a spin-wave-like ring (Figs. 2c and 2h). One-dimensional
cuts through Fig. 2d at $\hbar\omega=200\pm 15$ meV along 
four different directions (Figs. 1d-g) confirm
that the scattering is indeed isotropic and symmetric
around $(\pi,\pi)$ like spin waves. With increasing energy,
the ring continues to disperse outward until magnetic 
scattering is no longer discernible at $\hbar\omega=315\pm 15$ meV (Fig. 2j).

Figure 3a summarizes the dispersion of spin excitations determined from the cuts in Figs. 2f-j.
The dashed boxes show the positions of crystalline electric field (CEF) excitations arising from the Pr$^{3+}$ rare earth ions 
in the tetragonal structure of PLCCO \cite{boothroyd}. Compared to intensities of Pr$^{3+}$ CEF levels, 
Cu$^{2+}$ spin fluctuations in PLCCO are extremely weak and cannot be separated from the strong Pr$^{3+}$ scattering
at certain CEF energy positions.
For reference, figure 3b shows an energy cut along ${\bf Q}=(0.5,0.5,L)$ for the $E_i=115$ meV data in which it is clear that 
CEF intensities at $\hbar\omega\approx 20$, 85 meV are significant relative even to the incoherent elastic
scattering.

To estimate the strength of the magnetic exchange coupling, we consider a 
two-dimensional AF Heisenberg Hamiltonian with the nearest ($J_1$) and the next nearest 
($J_2$) neighbor coupling (Fig. 1c). Since the zone boundary spin fluctuations sensitive 
to $J_2$ were unobservable (Fig. 2j),
we set $J_2=0$ and determined that $J_1=162\pm 13$ meV renders the best fit to the data for $\hbar\omega\ge 100$ meV. The corresponding calculated one-magnon cross sections are plotted as the solid lines in Figs. 2h-2j, and the resulting dispersion relation is shown as the solid line in Fig. 3a.  
At high energies ($\hbar\omega\ge 100$ meV), 
the calculated spin wave dispersion coincides fairly well with the data, but the value of $J_1$
is considerably larger than the hole- (La$_2$CuO$_4$, $J_1=104$, $J_2=-18$ meV) \cite{coldea} and electron- 
(Pr$_2$CuO$_4$, $J_1=121$ meV) \cite{bourges97} doped parent compounds (Fig. 3a). 
Assuming as-grown insulating PLCCO has similar AF exchange coupling as Pr$_2$CuO$_4$ \cite{bourges97}, 
our data suggest that the high energy spin fluctuations in electron-doped PLCCO disperse 
faster than the spin waves of the insulating compound.  Therefore they are unlikely to arise from the weak static AF order in the material \cite{dai05}.

Although high energy spin excitations in PLCCO are spin-wave-like, 
the observed scattering for $\hbar\omega\le 80$ meV 
is substantially broader than those predicted by the linear spin-wave theory (Fig. 3a).
A cut at $\hbar\omega=8\pm 1$ meV through the $(\pi,\pi)$ point along
the $[\overline{1},1]$ direction confirms this point (Fig. 3c). For energies below 20 meV, the dispersion of spin fluctuations has a nearest neighbor coupling of $J_1=29\pm 2.5$ meV
(dash-dotted line in Fig. 3a). Therefore, the dispersion of PLCCO can be separated into two regimes.
For energies ($4\le\hbar\omega\le 80$ meV), the excitations are 
broad and weakly dispersive. For $\hbar\omega\ge 100$ meV, the fluctuations are spin-wave-like with $J_1$ larger than 
that of the insulating parent compound.

\begin{figure}
\includegraphics[scale=.33]{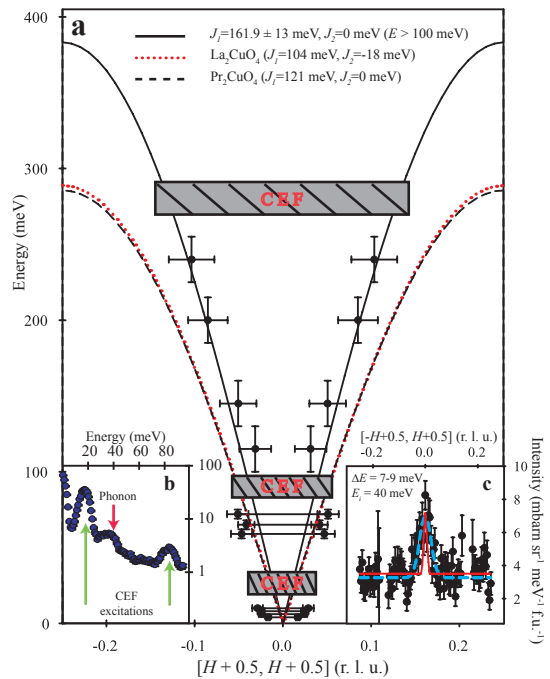}
\caption{a) The dispersion of spin excitations in PLCCO.  Points connected by solid lines denote 
only one continuous peak centered 
around $(\pi,\pi)$ and represent the FWHM of a Gaussian fit with the instrument resolution deconvoluted.  
Solid boxes show energy levels of Pr$^{3+}$ CEF scattering.
Solid, dotted, dashed, dash-dotted lines show dispersions 
from linear spin-wave fits with various exchange couplings.  
b) Log vs. linear energy-cut averaged from $H=0.45$ to 0.55 along the $[1,1]$ direction 
and from $H=-0.05$ to 0.05 along the $[\overline{1},1]$ direction. 
CEF and phonon contamination energies 
are marked by arrows.  c) ${\bf Q}$-cut along the $[\overline{1},1]$ direction with $\hbar\omega=8\pm1$ meV 
with a Gaussian fit in a blue dashed line and the calculated one-magnon cross section in solid red with $J$ = 121 meV.}
\end{figure}    

In addition to determining the dispersion of spin excitations in PLCCO, 
the absolute spin susceptibility $\chi^{\prime\prime}({\bf Q},\omega)$ measurements in Fig. 2 also allow us to calculate the energy dependence  
of the local susceptibility $\chi^{\prime\prime}(\omega)$, defined as $\int \chi^{\prime\prime}({\bf Q},\omega) d^3Q/\int d^3Q$ \cite{hayden96,dai99}.
Figure 4a shows how $\chi^{\prime\prime}(\omega)$ varies as a function of $\hbar\omega$ for electron-doped  superconducting PLCCO. Similar to hole-doped 
materials \cite{hayden96,tranquada,christensen,hayden,stock,stock1,dai99}, 
electron-doping suppresses the spectral weight of
spin fluctuations in PLCCO at high ($\ge 50$ meV) energies. For energies below 50 meV, $\chi^{\prime\prime}(\omega)$ increases with decreasing energy and does not
saturate at $\hbar\omega=4\pm 1$ meV, the lowest energy probed on MAPS. Assuming that the $\chi^{\prime\prime}(\omega)$ in crystals of
MAPS experiments ($T_c = 21\pm 1$ K) is similar to that of the previously studied $T_c=21$ K PLCCO sample \cite{dai05,kang05,wilson05}, we can normalize the low-energy magnetic response of the
$T_c=21$ K sample obtained on SPINS spectrometer at NIST to that of the MAPS data. The outcome, shown as inset of Fig. 4, reveals a new energy scale of 2-3 meV for superconducting PLCCO.

\begin{figure}
\includegraphics[scale=.33]{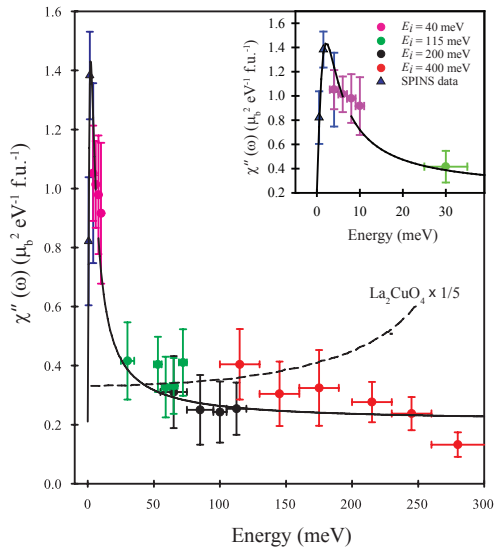}
\caption{Energy dependence of local susceptibility $\chi^{\prime\prime}(\omega)$ in PLCCO
determined from integration over wave vector of the observed magnetic
scattering around $(\pi,\pi)$ \cite{hayden96,dai99}. The dashed line shows
$\chi^{\prime\prime}(\omega)\times 1/5$ for La$_2$CuO$_4$ \cite{hayden96}.
Since ${\bf Q}$-cuts were made along the $[\overline{1},1]$ direction, 
the background scattering can be approximated by a constant.  The blue triangles indicate 
normalized cold neutron triple-axis data 
on a $T_c=21$ K PLCCO at energies below 3 meV obtained on the SPINS spectrometer at 
NIST center for neutron research \cite{wilson05}.  The inset
shows an expanded view of $\chi^{\prime\prime}(\omega)$ vs $\hbar\omega$ at low energies. The 
solid line is a damped Lorenztian on a constant background. }
\end{figure}
  
We are now in a position to compare the spin excitations of electron-doped PLCCO with that of the hole-doped LSCO \cite{tranquada,christensen} and 
YBCO \cite{bourges,hayden,stock,stock1,dai99}.  For hole-doped materials such as LSCO \cite{cheong,yamada98,tranquada,christensen} 
and YBCO with $x\ge 0.45$ \cite{dai,bourges,hayden,stock}, 
the low-energy spin fluctuations are incommensurate and display an inward dispersion 
toward a resonance point with increasing energy. This is not observed in electron-doped materials. 
Instead, spin fluctuations in PLCCO have a broad commensurate peak centered at $(\pi,\pi)$ at low-energies ($\le 50$ meV) which 
disperses outward into a continuous spin-wave, ring-like scattering at high energies ($\ge 100$ meV), similar to lightly doped YBCO with $x=0.353$ \cite{stock1}. 
At present, it is not clear how theoretical models based on spin stripes \cite{kivelson} can reconcile the
differences in spin excitations between the hole- and electron-doped materials. 

For hole-doped LSCO with  
$T_c=38.5$ K, the mean-squared fluctuating moment $\left\langle m^2\right\rangle=\int \chi^{\prime\prime}(\omega)d\omega$ integrated up 
to 40 meV is $\left\langle m^2\right\rangle=0.062\pm0.005$ $\mu_B^2$f.u.$^{-1}$ \cite{christensen}.
For comparison, $\left\langle m^2\right\rangle$ calculated from $\chi^{\prime\prime}(\omega)$ up to 40 meV is only $0.024\pm 0.003$  
$\mu_B^2$f.u.$^{-1}$ in PLCCO, about three times smaller than that of LSCO. The total fluctuating moment integrated 
from 0 to 300 meV (Fig. 4)
gives $\left\langle m^2\right\rangle=0.089\pm0.009$ $\mu_B^2$f.u.$^{-1}$, a value 
an order of magnitude larger than the static moment squared (0.0016 $\mu_B^2$f.u.$^{-1}$) \cite{dai05}.
Since the total moment sum rule for spin-$1\over2$ Heisenberg model requires one-magnon fluctuating
moment squared to be smaller than the ordered static moment squared \cite{lorenzana}, 
the observed high-energy spin-wave-like excitations are unlikely to arise from the small ordered moment.

In the standard Hubbard model and its strong-coupling limit, the $t$-$J$ model with
only the nearest-neighbor hopping $t$, there should be complete particle-hole 
symmetry and therefore the electron- and hole-doped copper oxides should behave 
identically. The observed large difference between incommensurate and commensurate 
spin fluctuations in hole- \cite{cheong,yamada98,tranquada,christensen} and electron-doped materials \cite{yamada99,yamada,dai05,kang05,wilson05}
has mostly been attributed to their differences 
in the strength of second nearest-neighbor ($t^\prime$) and third ($t^{\prime\prime}$)
nearest-neighbor hopping. This also
explains their differences in Fermi surface topology
 within the $t$-$J$ model \cite{li03,tohyama,yuan}, although it may also be due
to their proximity to two different quantum critical points \cite{onufrieva}. In the most recent
calculation using the slave-boson mean-field theory and random phase approximation \cite{yuan}, 
incommensurate spin fluctuations at $(0.3\pi,0.7\pi)$ have been predicted for optimally doped NCCO. 
However, this is not observed in our PLCCO (Figs. 2a-2e). Similarly, the energy dependence
of the $\chi^{\prime\prime}({\bf Q},\omega)$ at ${\bf Q}=(\pi,\pi)$ has been predicted to exhibit a peak
between $0.1\omega/J$ \cite{li03,yuan} and $0.4\omega/J$ \cite{tohyama}.  
While qualitatively similar to the predictions, $\chi^{\prime\prime}(\omega)$ in Fig. 4 has a peak at a much
smaller energy of $0.02\omega/J$. Comparison of future calculations in absolute units with our
data should determine whether itinerant magnetism models can account quantitatively for the observed 
dynamic susceptibility in superconducting PLCCO.

This work is supported by the U. S. NSF DMR-0453804 and DOE Nos. DE-FG02-05ER46202 and 
DE-AC05-00OR22725 with UT/Battelle LLC.


\end{document}